\documentclass{article}
\usepackage{spconf,amsmath,graphicx}

\usepackage{enumitem}
\setlist{nosep, leftmargin=14pt}

\usepackage{mwe} 
\usepackage{titlesec}
\usepackage{multirow}
\usepackage{booktabs}
\usepackage{colortbl}
\definecolor{lightgray}{gray}{0.9}
\usepackage{float}
\restylefloat{table}
\usepackage{stfloats}
\usepackage{url}

\title{Investigating Label Bias and Representational Sources of Age-Related Disparities in Medical Segmentation}

\name{Aditya Parikh \quad Sneha Das \quad Aasa Feragen}
\address{DTU Compute, Technical University of Denmark}
\begin{document}
%
\maketitle
\begin{abstract}

Algorithmic bias in medical imaging can perpetuate health disparities, yet its causes remain poorly understood in segmentation tasks. While fairness has been extensively studied in classification, segmentation remains underexplored despite its clinical importance. In breast cancer segmentation, models exhibit significant performance disparities against younger patients, commonly attributed to physiological differences in breast density. We audit the MAMA-MIA dataset, establishing a quantitative baseline of age-related bias in its automated labels, and reveal a critical Biased Ruler effect where systematically flawed labels for validation misrepresent a model's actual bias. However, whether this bias originates from lower-quality annotations (label bias) or from fundamentally more challenging image characteristics remains unclear. Through controlled experiments, we systematically refute hypotheses that the bias stems from label quality sensitivity or quantitative case difficulty imbalance. Balancing training data by difficulty fails to mitigate the disparity, revealing that younger patient cases are intrinsically harder to learn. We provide direct evidence that systemic bias is learned and amplified when training on biased, machine-generated labels, a critical finding for automated annotation pipelines. This work introduces a systematic framework for diagnosing algorithmic bias in medical segmentation and demonstrates that achieving fairness requires addressing qualitative distributional differences rather than merely balancing case counts.\footnote{Code for the experimental framework and evaluation is made available at \url{https://(to be released upon acceptance)}}

\end{abstract}
\vspace{-.2cm}
\section{Introduction}
\label{sec:intro}


The integration of deep learning in medical imaging has shown potential for automating critical tasks like tumor segmentation, yet it carries substantial risk of inheriting and amplifying biases present in clinical data \cite{xu2024addressing,STANLEY2025105501}. Algorithmic bias in healthcare is concerning and can perpetuate existing demographic disparities~\cite{flores2010racial}. In breast cancer diagnostics, a well-documented challenge is that segmentation performance is often lower for younger patients~\cite{kundu2025detecting,stephens2024race}. This is commonly attributed to their higher breast density~\cite{eisemann2025nationwide}, which can obscure tumor margins and complicate segmentation for {\it both} radiologists and automated systems. Meanwhile, fairness -- the principle that a model should not systematically disadvantage certain patient subgroups -- remains underexplored in segmentation, a task with direct implications for treatment planning and clinical decision-making. 


\begin{figure}[!t]
\begin{minipage}[b]{0.9\linewidth}
  \centering
  \centerline{\includegraphics[width=1.0\textwidth]{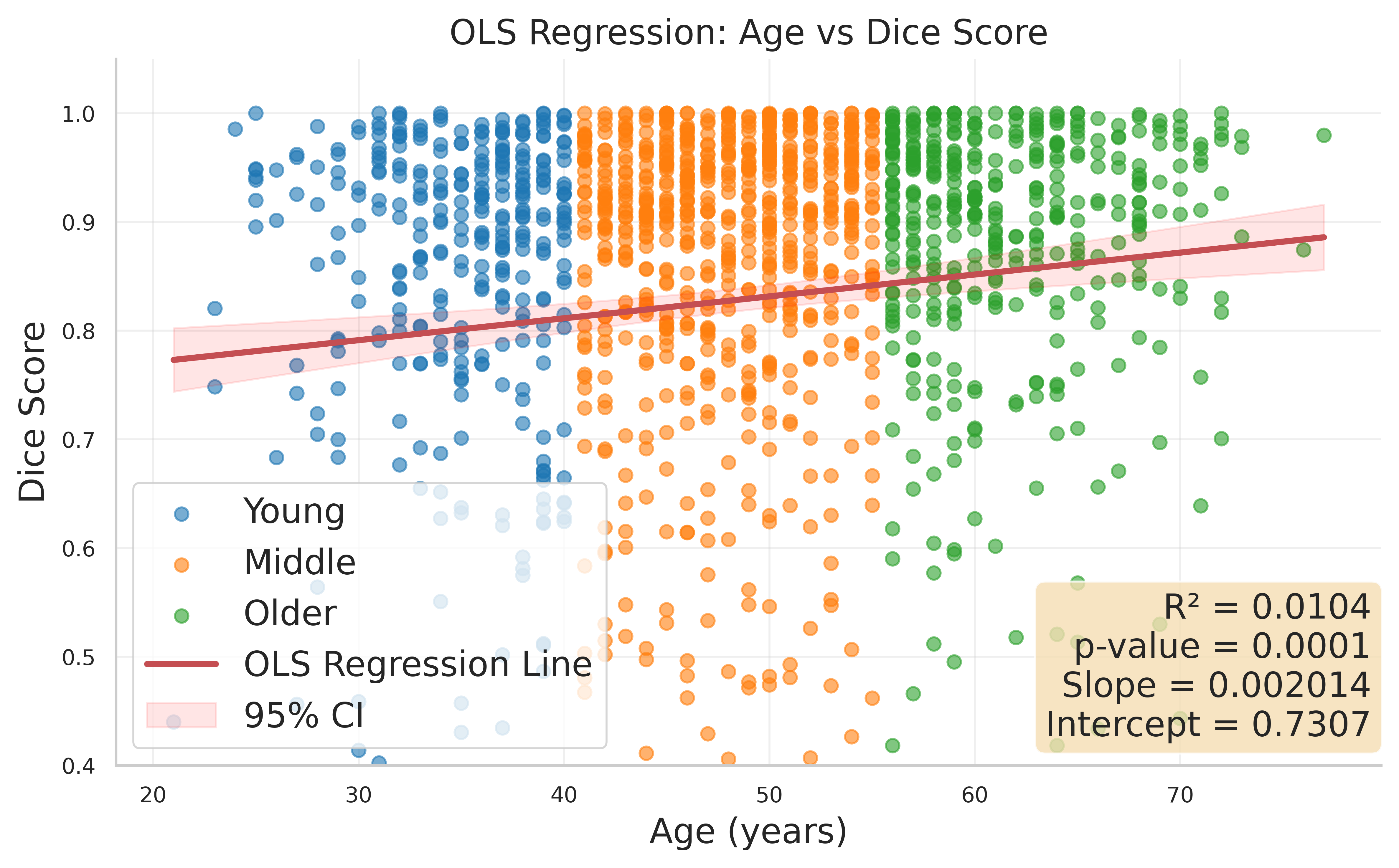}}
\end{minipage}
\caption{Inherent age-related bias in the Silver-Standard automated labels. An OLS regression reveals a positive correlation between patient age and segmentation performance (Dice Score), establishing a quantitative baseline of disparity.}
\label{fig:res}
\end{figure}

While the correlation between age and performance is established, its underlying reasons remain unclear. While higher segmentation difficulty due to higher breast density is a previously cited possible reason, there are other potential contributing factors. Performance disparities often come from \textbf{representational bias}, arising from differences in case distribution or intrinsic imaging characteristics across age groups \cite{tschida2025evaluating,kundu2025detecting}. This includes not only class imbalance and prevalence of challenging cases but also their qualitative nature. Factors like non-mass enhancement, irregular tumor morphology, or variable presentation patterns that may be more prevalent in younger women make them systematically harder for the model to learn~\cite{jamanetworkopen.2024.37402, woo2025invasive}. Another potential reason is \textbf{label bias} -- where the ground truth annotations for younger patients are systematically less accurate due to the inherent difficulty of manual segmentation. If the annotations have lower quality in younger subjects, then our ability to measure performance is also lower in the very same subjects.


Differentiating between the causes and sources of bias is critical for developing effective fairness interventions. To date, label bias has remained an unrecognised and unquantified source of bias in image segmentation. As segmentation benchmarks and even datasets used in real systems are often created using semi- or fully-automatic tools (e.g., ISIC Challenge datasets \cite{8363547}, FreeSurfer for neuroimaging~\cite{Hendrickson2023.03.22.533696}, nnU-Net-based annotations~\cite{isensee2021nnu, deepa_krishnaswamy_2023_7975081, garrucho2024mama}), we have every reason to assume that segmentation labels can have systematic biases. It is known from classification tasks that this gives a \emph{biased ruler} effect, where we are unable to effectively measure and mitigate bias. In this paper, we therefore let machine-generated labels from a pre-existing system serve as a methodological probe to measure a model's \textit{apparent performance} compared to its \textit{real} performance. This allows us to quantify the impact of using a flawed, real-world benchmark. 

In this paper, we thus present a systematic analysis designed to disentangle the potential sources of bias in image segmentation through a series of controlled experiments on the MAMA-MIA dataset \cite{garrucho2024mama}, a large-scale dataset of breast cancer MRI images. Our contributions are: (i) A first comprehensive \textbf{fairness audit of the MAMA-MIA dataset}, establishing a quantitative baseline of the age-related bias present in its automated labels; (ii) To the best of our knowledge, the \textbf{first study of label bias and its effect on bias audit in image segmentation}; (iii) A framework of controlled experiments designed to \textbf{isolate the effects of label bias from representational bias}; (iv) Direct quantitative evidence of bias amplification, demonstrating how \textbf{systemic bias is learned and propagated} through machine learning pipelines.

\section{Methodology}
\label{sec:methods}
\vspace{-.2cm}

\subsection{Dataset} 

The MAMA-MIA dataset \cite{garrucho2024mama} is a large, publicly available multi-center breast cancer benchmark of dynamic contrast-enhanced magnetic resonance images (DCE-MRI). It comprises 1,506 unique patient cases, each including volumetric imaging data, rich demographic metadata, and a pair of segmentation masks: An expert-annotated mask and a mask automatically generated by a standard nnU-Net framework~\cite{isensee2021nnu} trained on an external dataset, including resampling to a target isotropic voxel spacing of \([1.0\times1.0\times 1.0 mm]\) and Z-score intensity normalization. For our experiments, we utilize the second T1-weighted, post-contrast phase as 3D volumetric input for all models. The patient age was stratified into three distinct age cohorts: \textit{Young} ($\leq 40$, $n=349$), \textit{Middle} ($40-55$, $n=754$), and \textit{Older} ($\geq 55$, $n=400$), based on established clinical relevance in breast cancer diagnostics \cite{humlevik2025distinct}. \\
\noindent
{\bf Evaluation Benchmarks:} We use both types of masks as validation labels: The term \textbf{Gold-Standard Labels} is used to refer to expert-annotated labels, created by 16 expert radiologists. As these are likely the least affected by image quality biases, we regard these as ground truth, used as a definitive benchmark for measuring a model's \textit{true performance}. \textbf{Silver-Standard Labels} are the automated nnU-Net masks. {\it Note} that these resemble the semi- or fully-automatic annotations often found in real-world segmentation datasets, and are therefore a realistic assumption of what segmentation labels frequently look like. The silver-standard labels include dual-expert qualitative ratings (Good, Acceptable, Poor, Missed) assessing their visual quality. \\
\noindent
{\bf Case Difficulty Stratification:} 
To analyze label quality and case difficulty, we stratify cases into four tiers combining dual-expert ratings with silver-standard metrics (Dice, HD95) as summarized in Table \ref{tab:tier_definitions}. This approach provides a proxy that captures both clinical utility and geometric precision.

\noindent

\begin{table}[t!]
\centering
\resizebox{\columnwidth}{!}{%
\begin{tabular}{clc}
\toprule
\textbf{Tier} & \textbf{Criteria} & \textbf{Difficulty} \\ \toprule
\textit{Tier 1} & Both experts rate ``Good" AND & \\ 
(Unambiguously Good) & (Dice $\geq$ 0.80 \& HD95 $\leq$ 10) & \multirow{2}{*}{\textbf{Easy}} \\ \cmidrule{1-2} 
\textit{Tier 1.5} & Both experts rate ``Good" BUT & \\
(Expert-Metric Mismatch) & (Dice $<$ 0.80 OR HD95 $>$ 10) & \\ \midrule 
\textit{Tier 2} & Any case with ``Acceptable" & \\ 
(Clinically Acceptable) & ratings OR expert disagreement & \multirow{2}{*}{\textbf{Hard}} \\ \cmidrule{1-2} 
\textit{Tier 3} & Both experts rate ``Poor" & \\
(Unambiguously Poor) & & \\ \bottomrule
\end{tabular}%
\caption{Definition of Quality Tiers and Difficulty Categories}

\label{tab:tier_definitions}
}
\end{table}
\vspace{-.4cm}
\subsection{Experimental Framework}

{\bf Training Protocol:} All models are trained using the nnU-Net~\cite{isensee2021nnu} with \texttt{3d\_fullres} (3D Full Resolution) configuration for 1000 epochs with Adam optimizer and nnU-Net's standard data augmentation (random rotations, scaling, elastic deformations, and gamma transformations). All experiments use 5-fold age-stratified cross-validation with a fixed seed for generalizable and reproducible findings. \\
\noindent
{\bf Evaluation Metrics:} We validate the segmentation using Dice Score and 95th percentile Hausdorff Distance (HD95) for boundary accuracy. Demographic disparities are quantified via two complementary fairness metrics \cite{10.1145/3616865, PMID:35273279}: \textbf{Demographic Parity Difference (DPD)} measuring absolute performance gaps:
$\text{DPD} = |P(\hat{y}=1|A=a) - P(\hat{y}=1|A=b)|$,
while the \textbf{Disparate Impact Ratio (DIR)} captures relative disparities:
$\text{DIR} = \frac{\min(P(\hat{y}=1|A=a), P(\hat{y}=1|A=b))}{\max(P(\hat{y}=1|A=a), P(\hat{y}=1|A=b))}$.
Here, $\hat{y}=1$ represents the beneficial outcome (high-quality segmentation; here, Dice Score $>0.8$), $A$ denotes the sensitive attribute (age), and $a, b$ are distinct subgroups. DPD ranges from 0 (perfect parity) to 1 (maximum disparity), while DIR ranges from 0 to 1 (perfect fairness). Following the ``four-fifths rule" from \cite{equal1990uniform}, a DIR below 0.8 is commonly considered evidence of adverse impact. In our study, we specifically compute DPD$(Y|O)$ and DIR$(Y|O)$, where $Y$ and $O$ denote the \textit{Young} and \textit{Older} subgroups, respectively, as they represent the most extreme demographic contrast. Additionally, we report the \textbf{fairness gap} ($\S$) as the absolute difference in mean performance between the highest- and lowest-performing demographic subgroups \cite{10.1145/3701716.3715598}. Statistical significance of group differences was tested using OLS regression \cite{dismuke2006ordinary}~(\texttt{Performance \textasciitilde{} Age}) and ANOVA at $\alpha=0.05$.
\vspace{-.4cm}
\subsection{Controlled Experiments for Bias Source}

\begin{table*}[!t]
\centering
\caption{Experimental results present the mean performance (± standard deviation), the Fairness Gap, statistical significance of group differences (ANOVA p-value), and formal fairness metrics for each setting. \textsc{M-BASELINE} (Gold-Standard) serves as the primary reference for comparing the effects of different interventions. All evaluations are against the Gold-Standard ground truth, except where explicitly mentioned.}

\label{tab:exp1_exp2_exp3_exp4_combined}
\setlength{\tabcolsep}{4pt}
\renewcommand{\arraystretch}{0.95}
\resizebox{\textwidth}{!}{%
\begin{tabular}{lcc|cc|c|c}
\toprule
 & \multicolumn{2}{c|}{\textbf{Experiment 1}}
 & \multicolumn{2}{c|}{\textbf{Experiment 2}} 
 & \textbf{Experiment 3} 
 & \textbf{Experiment 4} \\
\cmidrule(lr){2-3} \cmidrule(lr){4-5} \cmidrule(lr){6-6} \cmidrule(lr){7-7}
\textbf{Age Group / Metric} &
\textbf{Observed Perf.} &
\cellcolor{lightgray}\textbf{True Perf.}$^{\dag}$ &
\textsc{M-Swap-Young} & \textsc{M-Swap-Older} &
\textsc{M-Diff-Bal} &
\textsc{M-Biased-Input} \\ \midrule
\quad Young (n=349)&
0.6941 $\pm$ 0.2489 &
\cellcolor{lightgray}0.7304 $\pm$ 0.2333 &
0.7320 $\pm$ 0.2381 & 0.7298 $\pm$ 0.2407 &
0.7317 $\pm$ 0.2314 &
0.6797 $\pm$ 0.2463 \\
\quad Middle (n=349)&
0.7104 $\pm$ 0.2399 &
\cellcolor{lightgray}0.7333 $\pm$ 0.2253 &
0.7379 $\pm$ 0.2193 & 0.7316 $\pm$ 0.2321 &
0.7308 $\pm$ 0.2220 &
0.7132 $\pm$ 0.2282 \\
\quad Older (n=349)&
0.7500 $\pm$ 0.2056 &
\cellcolor{lightgray}0.7703 $\pm$ 0.1899 &
0.7739 $\pm$ 0.1922 & 0.7869 $\pm$ 0.1755 &
0.7678 $\pm$ 0.1999 &
0.7458 $\pm$ 0.1991 \\ \midrule
\textbf{Average} &
\textbf{0.7182 $\pm$ 0.2334} &
\cellcolor{lightgray}\textbf{0.7446 $\pm$ 0.2178} &
\textbf{0.7479 $\pm$ 0.2182} & \textbf{0.7505 $\pm$ 0.2183} &
\textbf{0.7435 $\pm$ 0.2188} &
\textbf{0.7129 $\pm$ 0.2270} \\ \midrule
\textbf{Fairness Gap}~${\S}$ &
\textbf{0.0559} &
\cellcolor{lightgray}\textbf{0.0399} &
\textbf{0.0419} & \textbf{0.0571} &
\textbf{0.0361} &
\textbf{0.0661} \\
\textbf{ANOVA p-value} &
\textbf{0.0049**} &
\cellcolor{lightgray}\textbf{0.0260*} &
\textbf{0.0227*} & \textbf{0.0149*} &
\textbf{0.0481*} &
\textbf{0.0006**} \\ \midrule
\textbf{DPD $(Y|O)$} &
\textbf{0.1060} &
\cellcolor{lightgray}\textbf{0.0802} &
\textbf{0.0716} & \textbf{0.0777} &
\textbf{0.0762} &
\textbf{0.1146}$^{\ddag}$ \\
\textbf{DIR $(Y|O)$} &
\textbf{0.8150} &
\cellcolor{lightgray}\textbf{0.8710} &
\textbf{0.8853} & \textbf{0.8755} &
\textbf{0.8761} &
\textbf{0.7895}$^{\ddag}$ \\ \bottomrule
\multicolumn{7}{l}{\footnotesize{**$p < 0.01$, *$p < 0.05$ indicates statistically significant group differences.}} \\
\multicolumn{7}{l}{$^{\dag}$\footnotesize{Highlighted column indicates the \textsc{M-Baseline} (True Perf.) model, used as the reference for comparisons.}} \\
\multicolumn{7}{l}{$^{\ddag}$\footnotesize{Indicates a potential adverse impact, defined here by common heuristics: a DIR $< 0.80$ (the four-fifths rule).}} \\

\end{tabular}%
}
\end{table*}

\begin{figure}[b!]
\centering
\includegraphics[width=0.48\textwidth]{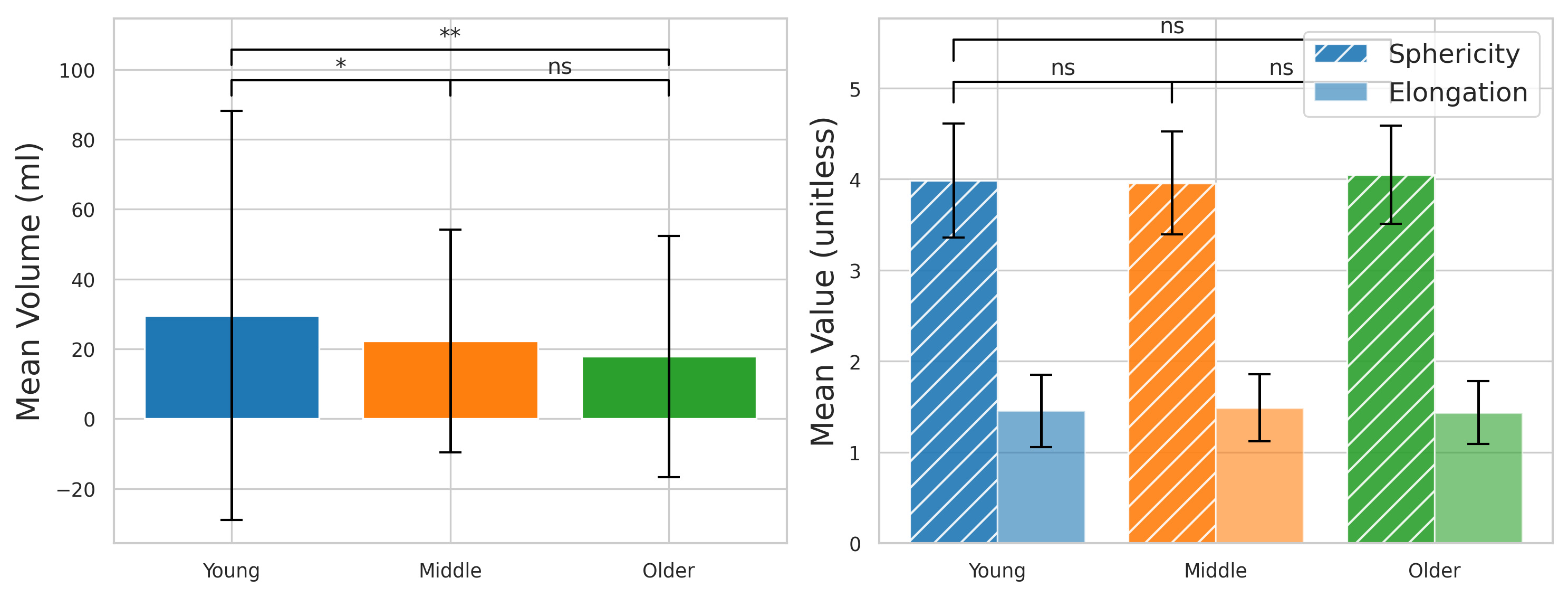}
\caption{(a) Tumor volume is, on average, larger and has higher variance in the \textit{Young} group ($p<0.01$ for Y-O). (b) In contrast, basic tumor shape metrics (sphericity and elongation) show no statistically significant difference.}
\label{fig:morphometrics}
\end{figure}

We conduct a sequence of controlled experiments to diagnose the reasons underlying bias: first we establish the existence of bias in the data and baseline model, then testing the hypothesis of label and representational bias, and finally demonstrating bias amplification when training on biased labels.\\
\noindent
\textbf{Experiment 0 - Establishing Anatomical Disparity and Benchmark Bias:} \label{exp:0} First, to test for an underlying anatomical basis of bias, we performed a morphometric analysis of Gold-Standard labels (see Fig. \ref{fig:morphometrics}). This reveals a {\it significant} disparity across age groups, where tumors in the \textit{Young} cohort are 66\% larger in volume and exhibit 70\% greater variance than \textit{Older} cohort. This provides evidence of an underlying anatomical {\it representational bias}. To further establish an initial baseline of real-world bias -- a fairness audit on the complete cohort of automated silver-standard labels reveals a significant relation between age and segmentation quality (OLS Regression; Dice score: $R^2=0.0104$, $p=0.0001$; HD95: $R^2=0.0093$, $p=0.0009$; see Fig. \ref{fig:res}). Formal fairness metrics support the finding (DPD: 0.0887; DIR: 0.699), indicating the \textit{Young} group achieves a high performance at only $\approx70\%$ the rate of the \textit{Older} group. {\it This confirms that the Silver-Standard labels are indeed systematically biased, motivating our subsequent experiments.}\\
\noindent
\textbf{Experiment 1 - The ``Biased Ruler" Effect:} Having confirmed a bias in the Silver-Standard labels, we next investigate the effect of using biased labels for validation, here exemplified by the Silver-Standard labels. This is particularly relevant as many segmentation datasets, both public benchmarks and those used for development of real-life (bio)medical imaging segmentation tools, are based on semi-automatic "ground truth" annotations that could be similarly biased. We use the Silver-Standard labels as an example of \emph{observed} labels, and compare performance with respect to biased observed labels to the true performance quantified using the Gold-Standard labels, in this experiment considered \emph{true} labels. 

To this end, we train an \textsc{M-Baseline} model under an age class-balanced protocol (n = 349 per group). The results (Table \ref{tab:exp1_exp2_exp3_exp4_combined}) show a statistically significant \emph{observed} performance bias against the \textit{Young} cohort $({\S}=0.0559$, $p=0.0049)$ when using the observed labels, which is $40\%$ higher than the \emph{true} bias $({\S} = 0.0399,~p=0.0260)$. This inflation of observed bias is also reflected in the formal fairness metrics (DPD: $0.0802 \rightarrow 0.1060$; DIR: $0.8710 \rightarrow 0.8150$). {\it This ``Biased Ruler" effect quantitatively demonstrates how relying on flawed benchmarks for fairness auditing can misrepresent the model's true performance disparities.}\\
\noindent
\textbf{Experiment 2 - Label Bias Sensitivity:} We now investigate whether the true bias in \textsc{M-Baseline} was caused by a superficial sensitivity to label quality. The \textsc{M-Swap-Young} and \textsc{M-Swap-Older} replaced $100\%$ of the \textit{Tier 1} labels with their biased, Silver-Standard counterparts. This intervention had no meaningful effect on the model's bias, see Table \ref{tab:exp1_exp2_exp3_exp4_combined}. The fairness gap remained stable, {\it refuting the hypothesis that the models' fairness is fragile or primarily driven by a small subset of high-quality labels for any specific subgroup.} \\
\noindent
\textbf{Experiment 3 - Persistence of Representational Bias:} Here, we test whether the true bias was caused by a quantitative imbalance in the distribution of \textit{hard} cases. The \textsc{M-Diff-Bal} model is trained on a dataset carefully balanced to provide each age group with an identical distribution of \textit{easy} $(n=143)$ and \textit{hard} $(n=206)$ cases. The results (Table \ref{tab:exp1_exp2_exp3_exp4_combined}) show that the difficulty-balancing intervention failed to eliminate the bias, leaving the fairness gap unchanged. This {\it refutes the hypothesis that a simple quantitative imbalance of difficult cases causes the bias.} \\
\noindent
\textbf{Experiment 4 - Training on Biased Labels Amplifies Bias:} In this experiment, we train an age-group balanced model \textsc{M-Biased-Input} on biased Silver-Standard labels. The results (in Table \ref{tab:exp1_exp2_exp3_exp4_combined}) confirm the hypothesis of \textbf{bias amplification}. The fairness gap widened by $66\%$ relative to \textsc{M-Baseline} (${\S}$ from $0.0399 \rightarrow 0.0661$), and the bias became statistically severe $(p=0.0006)$. This amplification is even more evident when viewed through formal fairness metrics: the DIR dropped below the standard threshold to $0.7895$. 
\vspace{-.2cm}
\section{Results and Conclusion}

Our experimental study systematically diagnosed the mechanism behind age bias in breast cancer tumor segmentation. We first demonstrate (Experiment 0) that automated Silver-Standard segmentation labels are biased, and that the age-balanced model exhibits a statistically significant bias whose observed magnitude would be substantially inflated (40\%) if we had only observed performance with respect to the automated Silver-Standard labels (Experiment 1). This effect has critical clinical implications: In real-world deployment, such a biased evaluation framework could mask true model performance, leading to undetected diagnostic failures and delayed treatment interventions. Moreover, if biased segmentation labels are used to validate model updates and guide clinical thresholds, this may systematically disadvantage younger patients by setting performance standards that appear adequate but actually mask age-related disparities. This underscores the urgent need for awareness of the effects of label bias in segmentation validation.

\begin{figure}[!t]
\centering
\begin{minipage}[b]{.48\linewidth}
  \centering
  \centerline{\includegraphics[width=\linewidth]{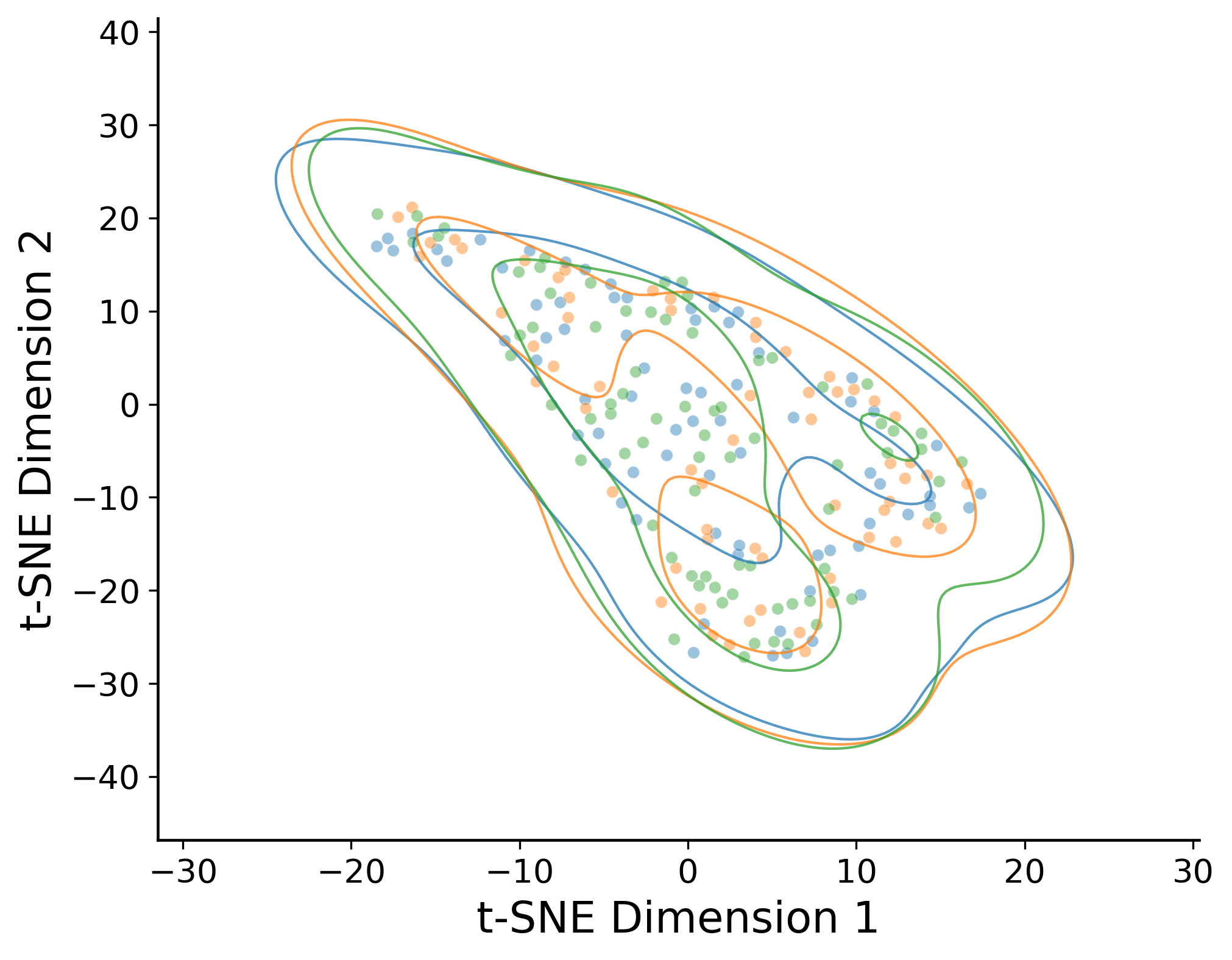}}
\end{minipage}
\hfill
\begin{minipage}[b]{.50\linewidth}
  \centering
  \centerline{\includegraphics[width=\linewidth]{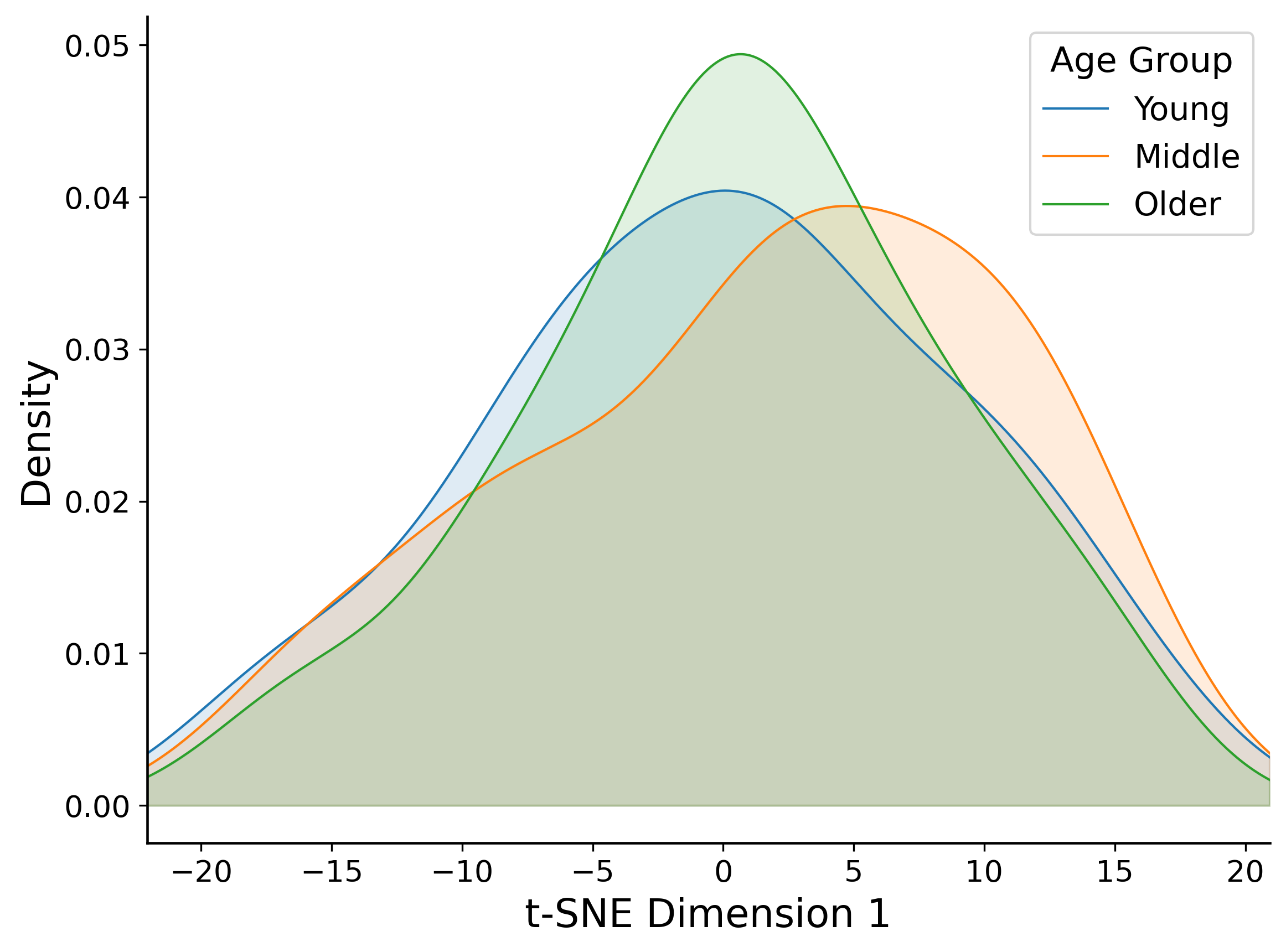}}
\end{minipage}

\caption{\textbf{Inspection of subgroup distribution shifts in the feature space projection.} t-SNE (t-distributed stochastic neighbor embedding) \cite{maaten2008visualizing} embeddings (for representative fold 0). \textit{Left:} Scatter plot of the first two t-SNE dimensions. \textit{Right:} Density distribution of the first t-SNE dimension. Both plots indicate a strong overlap, suggesting the model’s latent space does not strongly separate representations by age. Clustering metrics quantify this overlap across 5-folds (Mean$\pm$Std): For t-SNE, Silhouette = \textbf{$-0.0312 \pm 0.0112$}, Purity = \textbf{$0.3943 \pm 0.0193$}, ARI (Adjusted Rand Index; [-1, 1], higher is better agreement) = \textbf{$0.0033 \pm 0.0113$}, NMI (Normalized Mutual Information; [0, 1], higher is better agreement) = \textbf{$0.0126 \pm 0.0098$}. Low ARI and NMI values confirm poor correspondence between the embedding structure and true age groups. t-SNE parameters: perplexity~$\approx n/10$ (clipped to~5–50), LR~200, iters~1500, init~=~PCA.}
\vspace{-.5cm}
\label{fig:embedding_fold0}
\end{figure}

The following experiments (Experiments 2-3) refute the hypothesis that bias stems from superficial sensitivity to label quality, where replacing high-quality labels with biased counterparts had no meaningful effect on fairness gaps. The difficulty-balancing intervention then isolated representation as a definitive cause. The failure to completely eliminate bias point out that the problem is not the \textit{quantity} of hard cases, but their qualitative nature. Our morphometric analysis provides supporting evidence: \textit{Hard} cases for \textit{Young} group are part of an anatomically distinct distribution. This aligns with clinical literature reporting that breast cancers in younger women are often larger and more aggressive~\cite{MAGGARD2003109, albasri2021clinicopathological}, and physiological factors mentioned in Section~\ref{sec:intro}. Visual analysis of learned representations supports this conclusion (see Fig.~\ref{fig:embedding_fold0}).

More critically, training on biased Silver-Standard labels widened the fairness gap by $66\%$ (Experiment 4). This creates dangerous clinical implications for modern AI development pipelines, where models are increasingly retrained on machine-generated labels to scale annotation efforts. For younger patients, this amplification may degrade segmentation quality and affect treatment planning. 

In conclusion, we provide both a systematic diagnosis of age-related algorithmic bias and a demonstration of its serious clinical implications. The nature of this bias is learnable, amplified by label bias, and rooted in qualitative representational disparities, which demand fundamental changes to fairness practices in medical imaging. Future work must address qualitative representational interventions rather than rebalancing strategies. Additionally, rigorous auditing protocols using high-quality benchmarks must be established to detect and prevent bias propagation in automated data pipelines, ensuring that efforts to scale AI systems do not unintentionally scale their inequities.

\section{Acknowledgment}

This work was funded by the Novo Nordisk Foundation under project number 0087102.

\bibliographystyle{IEEEbib}
\bibliography{refs}

\end{document}